\documentclass[aps,twocolumn,superscriptaddress,showpacs]{revtex4}

\usepackage{amsmath}
\usepackage{amssymb}
\usepackage{amsfonts}
\usepackage{graphicx}
\usepackage{bm}

\usepackage{dcolumn}
\usepackage{bm}
\usepackage{graphicx}
\usepackage{color}
\DeclareGraphicsExtensions{. jpg,. pdf, . mps, . png, . eps, . ps, . EPS}

\begin{document}
\def\be{\begin{equation}}
\def\ee{\end{equation}}

\def\bc{\begin{center}} 
\def\ec{\end{center}}
\def\bea{\begin{eqnarray}}
\def\eea{\end{eqnarray}}
\newcommand{\avg}[1]{\langle{#1}\rangle}
\newcommand{\Avg}[1]{\left\langle{#1}\right\rangle}

\title{Quantum mechanical formalism for biological evolution }

\author{Ginestra Bianconi} 
\affiliation{Department of Physics, Northeastern University, Boston, 
Massachusetts 02115 USA}
\author{Christoph Rahmede}
\affiliation{Department of Physics and Astronomy, University of Sussex, Brighton,BN1 9QH,UK}

\begin{abstract}
We study the evolution of sexual and asexual populations in general fitness landscapes.
We find deep relations between the mathematics of biological evolution and the formalism of quantum mechanics.
We give the general structure of the evolution of populations which is in general an off-equilibrium process that can be expressed by path integrals over phylogenies. These phylogenies are sums of linear lineages for asexual populations.  For sexual populations instead,  each lineage is a tree of branching ratio two  and the path integral describing the evolving population is given by a sum over these trees.
Finally,  we show that  the Bose-Einstein and the Fermi-Dirac distributions  describe the stationary state of  biological populations in simple cases.
\end{abstract}

\maketitle
  \section{Introduction}

The intriguing relation   between evolutionary dynamics
and statistical mechanics \cite{Fisher,Kimura, Hirsh, Gerland} has attracted the interest of great evolutionary theorists like Fisher \cite{Fisher} or Kimura  \cite{Kimura} that have related their results to the second principle of thermodynamics or to the theory of gases.
Interestingly, the relation between evolutionary theory and quantum statistical mechanics is emerging  from  a series of independent works \cite{Kingman,Krug,Bose,Fermi, Complex,W,Kadanoff,Ferretti,Shraiman} that show a class of  phase transitions occurring in evolution of haploid populations and other evolving complex systems described  by a   Bose-Einstein condensation. In haploid populations, this transition is the quasi-species phase transition   \cite{Eigen,Nowak,Sigmund,Gillespie, Hartl,Sequence} in which a finite fraction of an asexual population ends up having the same genotype if the selective pressure is over a  critical value and the mutation rate is smaller than a critical value.
Moreover,  in  a recent paper \cite{Bose_n} it was shown that a  condensation transition in the Bose-Einstein universality class   occurs also  in the evolution  of diploid sexual populations in presence of epistatic interactions. When this condensation occurs, a finite fraction of pairs of genetic loci in epistatic interactions is fixed.
 
 The deep relation between evolutionary theory and the formalism of quantum mechanics extends also to the dynamical description of  biological  evolution. The quantum spin-chain formalism has been shown to  solve models of asexual evolution   \cite{Baake}.
Moreover, two  papers have recently highlighted in specific cases (asexual evolution in mean-field landscapes   \cite{Peliti} and asexual evolution in fluctuating environment   \cite{Leibler}) the role of path integrals in describing the temporal evolution of populations.

In this paper we propose a general theoretical framework that  reconciles all the cited  results and extends both to asexual and sexual populations defined over complex fitness landscapes. The analogies that are present  at mathematical level between biological evolution and quantum mechanics are highlighted.
In particular  we explore the  different  nature of the path integral calculation that solves for the dynamical evolution of complex populations.
 The difference that resides between asexual and sexual evolution is that these phylogenies are sums of linear lineages for asexual populations, instead for sexual populations they are sums over trees of ancestors, each individual having two parents, four grandparents and so on. 
Finally we show  that quantum statistics (both Fermi-Dirac and Bose-Einstein statistics) characterize the steady state distribution of sexual and asexual populations in simple cases.  
 
 The paper is organized as follows: In section II we present the equations of evolution of asexual populations and in section III we present the  equations and the structure of their solution for sexual populations.
In each of these sections we first study the off-equilibrium evolution with non-overlapping generations, then we study the evolution with overlapping generations and finally we characterize the steady state distribution of the evolutionary dynamics in simple cases.
Finally in section IV we give the conclusions.

\section{Evolution of asexual populations}

The genome of an asexual organism is formed by a single copy of each chromosome.
If we indicate by $i=1,\ldots , N $ a genetic locus, a given genotype is determined by
the allelic states $\{\sigma\}=(\sigma_1,\sigma_2,\ldots,\sigma_i,\ldots \sigma_N)$ at each genetic locus $i$.
 The allelic state $\sigma_i$ at each genetic locus  $i$ can take 4 values corresponding to  adenine, thymine, cytosine, guanine, i.e.  $ \sigma_i=1,2,3,4$.
The  population evolves under the drive of selection that favors allelic configurations corresponding to higher reproduction rate,  and mutations that increase the genetic variation in the population.
We assume that the reproductive rate $W(\{\sigma\})$, also called Wright fitness, of a genotype $\{\sigma\}$ is given by 
\begin{equation}W(\{\sigma\})=e^{-\beta U(\{\sigma\})}\end{equation} where $U(\{\sigma\})$ is the Fisher fitness and $\beta$ is   the {\em selective pressure}. If $\beta=0$ every genotype has the same reproductive rate. If $\beta\gg1$ the difference in reproductive rate of  genotypes having different $U(\{\sigma\})$ is strongly enhanced.

\subsection{Non-overlapping generations}

In the literature, usually the case of non-overlapping generations is studied   \cite{Peliti,Sigmund}. The equations for asexual evolution are written in terms of the probability $\tilde{P}^t(\{\sigma\})$ that at generation $t$ an individual of the population has genotype $\{\sigma\}$.
The equation for the evolutionary dynamics in non-overlapping generations is  given by   \cite{Nowak,Krug,Peliti}
\begin{equation}
\tilde{P}^{t+1}(\{\sigma\})=\frac{\sum_{\{\sigma^{\prime}\}}Q(\{\sigma\}|\{\sigma^{\prime}\})e^{-\beta U(\{\sigma^{\prime}\})}\tilde{P}^t(\{\sigma^{\prime}\})}{Z_t}
\label{pno}
\end{equation}
where the normalization constant $Z_t$ is given by 
\begin{equation}
Z_t=\sum_{\{\sigma\}}\sum_{\{\sigma^{\prime}\}}Q(\{\sigma\}|\{\sigma^{\prime}\})e^{-\beta U(\{\sigma^{\prime}\})}\tilde{P}^t(\{\sigma^{\prime}\})\ .
\end{equation}
The matrix $Q(\{\sigma\}|\{\sigma^{\prime}\})$  in Eq. $(\ref{pno})$ represents the probability of mutations between genotype $\{\sigma^{\prime}\}$ to genotypes $\{\sigma\}$  in the next generation. 
Therefore, if $\mu$ is the mutation rate  we have 
\begin{equation}
Q(\{\sigma\}|\{\sigma^{\prime}\})=\prod_i\left[(1-\mu)\delta(\sigma_i, \sigma^{\prime}_i)+\frac{\mu}{4}\right]\ .
\label{Q1}
\end{equation}

The solution of this  evolutionary dynamics is given by a path integral over the phylogenies of the population   \cite{Baake,Peliti}. In fact if we iterate the Eq. $(\ref{pno})$ we find the solution 
\begin{widetext}\begin{equation}
\tilde{P}^{t}(\{\sigma\})=\frac{1}{{{\cal Z}_t}}\sum_{\{{\sigma}\}_{t-1}} \sum_{\{\sigma\}_{t-2}}\ldots \sum_{\{\sigma\}_{0}}\prod_{\tau=0}^{t-1} Q(\{\sigma\}_{\tau+1}|\{\sigma\}_{\tau})e^{-\beta U(\{\sigma\}_{\tau})}\tilde{P}^0(\{\sigma\}_{0})\ ,
\label{pi}
\end{equation}
\end{widetext}
where the normalization constant is given by 
\begin{widetext}\begin{equation}
{{{\cal Z}_t}}=\prod_{\tau=0}^{t}Z_{\tau}=\sum_{\{\sigma\}_{t}} \sum_{\{\sigma\}_{t-1}}\ldots \sum_{\{\sigma\}_{0}}\prod_{\tau=0}^{t} Q(\{\sigma\}_{\tau+1}|\{\sigma\}_{\tau})e^{-\beta U(\{\sigma\}_{\tau})}\tilde{P}^0(\{\sigma\}_{0})\ .
\end{equation}
\end{widetext}
The  path integral in Eq. ($\ref{pi}$) can be directly calculated if we  solve the  eigenvalue problem 
\begin{equation}
{\mathbf M}_{\{\sigma\}|\{\sigma^{\prime}\}}\left[e^{-\beta U(\{\sigma^{\prime}\})}\pi_n(\{\sigma^{\prime}\})\right]=\lambda_n \pi_n(\{\sigma\})
\label{eigp1}
\end{equation}
where the  
 operator 
 ${\mathbf M}_{\{\sigma\}|\{\sigma^{\prime}\}}$ acts on a generic function $f(\{\sigma^{\prime}\})$
according to the rule
\begin{equation}
{\mathbf M}_{\{\sigma\}|\{\sigma^{\prime}\}} f(\{\sigma^{\prime}\})=\sum_{\{\sigma^{\prime}\}}Q(\{\sigma\}|\{\sigma^{\prime}\})f(\{\sigma^{\prime}\})\ .
\label{M1}
\end{equation}
The eigenfunctions $\pi_n(\{\sigma\})$ of the eigenvalue problem $(\ref{eigp1})$  are chosen to be normalized.
Let us  decompose the  probability distribution $\tilde{P}^t\{\sigma\})$ on the basis of the eigenfunctions $\pi_n(\{\sigma\})$ according to the following expression, i.e. 
\begin{equation}
\tilde{P}^t(\{\sigma\})=\sum_n d_n(t) \pi_n(\{\sigma\}).
\end{equation}
Using $(\ref{pno})$ we obtain the dynamical equation for the coefficients $d_n(t)$, i.e.
\begin{equation}
d_n(t)=d_n(0)\lambda_n^t.
\end{equation}
Therefore, asymptotically in time  the population is converging to the distribution corresponding to the maximal eigenvalue $\lambda_0=\max_n\lambda_n$.
The partition function ${\cal Z}_t$ is then given by
\begin{equation}
{\cal Z}_t=\sum_n d_n(0) \lambda_n^t.
\end{equation}

\subsection{Overlapping generations}
We now write the equation of biological evolution in continuous time.
Therefore we assume that at each time there can be a birth or a death process. We assume that the birth process depends on the reproductive rate $W(\{\sigma\})=\exp[-\beta U(\{\sigma\})]$ and that the death process is a random drift.
If we define the probability $P(\{\sigma\},t)$ that at time $t$ and individual has genome $\{\sigma\}$,
the dynamic equation of evolution of $P(\{\sigma\},t)$ is  given by
\be
\frac{d P(\{\sigma\},t)}{dt}={\mathbf M}_{\{\sigma\}|\{\sigma^{\prime}\}}\left[\frac{e^{-\beta U(\{\sigma^{\prime}\})}P(\{\sigma^{\prime}\},t)}{Z_t}\right]-P(\{\sigma\},t)
 \label{ev}
\ee
where  the operator ${\mathbf M}_{\{\sigma\}|\{\sigma^{\prime}\}}$ is defined in $(\ref{M1})$ and $Q(\{\sigma\}|\{\sigma^{\prime}\})$ is given by $(\ref{Q1})$.
 The partition function $Z_t$ in Eq. $(\ref{ev})$ is given by
\begin{equation}
Z_t=\sum_{\{\sigma\}}\sum_{\{\sigma^{\prime}\}}Q({\{\sigma\}|\{\sigma^{\prime}\}})e^{-\beta U(\{\sigma^{\prime}\})}P(\{\sigma^{\prime}\},t).
\label{defZ2}
\end{equation}
We observe here that the  equation $(\ref{ev})$ can be reduced  to the well known quasi-species  equation  \cite{Nowak,Sigmund,Gillespie, Hartl,Sequence} if we make a change of variables $t\to t^{\prime}$ with $dt/Z_t=dt^{\prime}$.
Let us now assume to know the solution of the eigenvalue problem $(\ref{eigp1})$
with $\lambda_n$ describing the discrete spectrum of this problem and $\pi_n(\{x\})$ the normalized eigenfunctions.
If we decompose the function $P(\{\sigma\},t )$ on the basis $\pi_n(\{\sigma\})$ of eigenfunctions , i.e.
\begin{equation}
P(\{\sigma\},t)=\sum_n c_n(t) \pi_n(\{\sigma\})\ ,
\end{equation}
the dynamical solution of Eq. $(\ref{ev})$ for the coefficients $c_n(t)$ is given by 
\begin{equation}
c_n(t)=\exp\left[ {\lambda_n}G(t)-t\right]c_n(0)\ .
\label{dyn2}
\end{equation}
In Eq. $(\ref{dyn2})$ the function $G(t)$ is defined through the function $Z_t$ according to the equation
\begin{eqnarray}
G(t)=\int_0^t dt^{\prime}\frac{1}{Z_{t^{\prime}}} \ .
\label{G}
\end{eqnarray}
Using the definition for $Z_t$ given by Eq. $(\ref{defZ2})$ together with Eqs. $(\ref{Q1})$  and $(\ref{dyn2})$, we can  close the self-consistent equations and  uniquely  determine the evolutionary dynamics of the population. Therefore the partition function $Z_t$ is given by
\begin{eqnarray}
Z_t=\sum_n \lambda^n c_n(t)=\avg{\lambda}.
\end{eqnarray}
where the average $\avg{\cdot}$ is performed over the functions $c_n(t)$ given by $(\ref{dyn2})$.
Finally, using $(\ref{dyn2})$ we can derive the equation obeyed by the  partition function $Z_t=\avg{\lambda}$,
\begin{equation}
\frac{1}{2}\frac{d\avg{\lambda}^2}{dt}=(\avg{\lambda^2}-\avg{\lambda}^2)\geq 0.
\label{FD}
\end{equation}
This equation is related to the Fisher theorem of natural selection   \cite{Fisher,Leibler} and describes the fact that evolution is an off-equilibrium process.
In fact, since $\avg{\lambda}=\Avg{W(\{\sigma\})}$ is the average reproductive rate of the population, Eq. $(\ref{FD})$ expresses the fact that this average reproductive rate always increases asymptotically in time as long as the environment does not change, i.e.  the Fisher fitness function $U(\{\sigma\})$ and the selective pressure remain constant in time. 
Eq. $(\ref{ev})$ shows surprising similarities with quantum mechanics that have been highlighted in a recent  paper    \cite{particelle} presenting  a unified framework  between this evolutionary equation and stochastic quantization.

\subsection{ The Bose-Einstein condensation in the  Kingman model }

The Kingman model   \cite{Kingman,Krug} is one of the most interesting stylized models of asexual evolution where the quasi-species transition is observed.
In the framework of this model the quasi-species transition can be exactly mapped to the Bose-Einstein condensation in a Bose gas.
One interesting aspect of the Kingman model is that the evolutionary dynamics reaches an equilibrium due to the constant drive of random mutations. 
In the Kingman model each individual is assigned a single real  parameter $\epsilon\geq 0$ determining its reproductive rate $W(\epsilon)$, i.e. 
\begin{equation}
W(\epsilon)=e^{-\beta \epsilon}\ .
\end{equation}
Moreover, in this model, when a  mutation occurs a new offspring is generated  with random fitness $\epsilon$ drawn  from a given distribution $\rho(\epsilon)$.
Therefore, instead of writing Eq. $(\ref{ev})$ for the probability $P(\{\sigma\})$ that an individual of the population has genotype $\{\sigma\}$, we can write the equation for the probability ${\cal P}(\epsilon,t |t_0)$ that a random individual in the population is associated with a  given Fisher fitness  $\epsilon$ at time $t$ and had the last mutation at time $t_0$.
The evolution of of $P(\epsilon,t |t_0)$ is given by 
\be
\frac{d P(\epsilon,t|t_0)}{dt}=\frac{e^{-\beta \epsilon^{\prime}}P(\epsilon,t|t_0)}{Z_t}-P(\epsilon,t|t_0)\ .
 \label{evB}
\ee
The partition function $Z_t$ in Eq. (\ref{evB}) is given by 
\be
Z_t=\int_0^t dt_0 \int d\epsilon\, e^{-\beta \epsilon}P(\epsilon,t|t_0)\ .
\label{defZB}
\ee
Therefore, the   probability $P(\epsilon,t|t_0)$ that an individual has  fitness $\epsilon$ at time $t$ under the condition that the last  mutation happened at time $t_0$, is given by the solution of $(\ref{evB})$, i.e.
\begin{equation}
P(\epsilon,t|t_0)=\alpha\, \rho(\epsilon)\, e^{\int_{t_0}^t dt\exp[-\beta \epsilon] (1-\alpha)\frac{1}{Z_t}-(t-t_0)}\ .
\end{equation}
Finally, integrating over $t_0$, we can evaluate the  probability $P(\epsilon,t)$ that an individual  at time $t$ has a given Fisher fitness  $\epsilon$ independently of $t_0$, i.e.
\begin{equation}
P(\epsilon,t)=\int_0^t dt_0\alpha \rho(\epsilon) e^{\int_{t_0}^t dt\exp[-\beta \epsilon] (1-\alpha)\frac{1}{Z_t}-(t-t_0)}.\nonumber
\end{equation}
In the case in which the distribution of the fitness has a finite support, one can easily find that he partition function $Z_t$ converges to a time independent  value $Z$.
Therefore the  
steady state solution for ${P}^{\star}_B (\epsilon)$  reached in the limit $t\to\infty$   is given by 
\begin{equation}
{P}^{\star}_B(\epsilon)=\mu\,\rho(\epsilon)\left[1+\frac{1}{{e^{\beta (\epsilon-\mu_B)}}-1}\right].
\label{BE}
\end{equation}
where $e^{-\beta \mu_B}=\Avg{e^{-\beta \epsilon}}/(1-\mu)$.
Therefore the  probability that an individual has fitness $\epsilon$ is determined by the Bose-Einstein distribution.
The probability distribution $P^{\star}_B(\epsilon)$ must satisfy the normalization condition
\begin{equation}
1=\int d\epsilon P^{\star}_B(\epsilon)=\mu\left[1+\int d\epsilon \rho(\epsilon)\frac{1}{{e^{\beta (\epsilon-\mu_B)}}-1}\right].
\label{BE2}
\end{equation}
When $\rho(\epsilon)$ is vanishing for $\epsilon\to 0$, the Bose-Einstein integral in Eq. $(\ref{BE2})$ can be limited from above. As a result, at  high enough selection pressure and low enough mutation rate the system might undergo a condensation phase transition in the Bose-Einstein universality class. 
Below this phase transition a finite fraction of the individuals in the population shares the same  genotype corresponding to  the maximal fitness.
This is one of the principal examples that show the so called quasi-species transition: for low mutation rate and high selective pressure a finite fraction of the population is found to have the same genotype.
Interestingly a similar phase transition occurs also in evolving ecologies   \cite{Ferretti}  where the invasive species might strongly reduce the biodiversity, and in  evolving models of  complex networks   \cite{Bose,W} where there might be the emergence of super-hubs like Google in the World-Wide-Web.

\subsection{The Fermi-Dirac  distribution in presence of negative selection}
The Fermi-Dirac distribution   can be obtained in presence of negative selection by a similar mechanism that generates the Bose-Einstein distribution in the Kingman model.
This model is mostly interesting because of the underlying symmetry between Fermi-Dirac and Bose-Einstein distribution.
We assume that each individual is assigned a parameter  $\epsilon$ describing its adaptability to the environment. The birth probability is one, and new individuals are generated at each time with parameter $\epsilon$ drawn by a given distribution $\rho(\epsilon)$. The  death probability  is given by a random drift with probability $\alpha$ and by a negative selection proportional to $\exp[\beta \epsilon]$ with probability $1-\alpha$.  
Therefore  the dynamics of evolution for the probability $P(\epsilon,t|t_0)$ that in the population there is an individual with parameter  $\epsilon$ born at time $t_0$ is given by 
\be
\frac{d P(\epsilon,t|t_0)}{dt}=-\left[(1-\alpha) \frac{e^{\beta \epsilon}}{Z_t}+\alpha\right]P(\epsilon,t|t_0)\ .
 \label{evF}
\ee
The partition function $Z_t$ in Eq. (\ref{evF}) is given by 
\be
Z_t=\int_0^t dt_0 \int d\epsilon e^{\beta \epsilon}P(\epsilon,t|t_0)\ .
\label{defZF}
\ee
Solving Eq. $(\ref{evF})$ we find the probability that an individual has fitness $\epsilon$ at time $t$ under the condition that it was born at time  $t_0$,
\begin{equation}
P(\epsilon,t|t_0)= \rho(\epsilon) e^{\int_{t_0}^t dt\exp[\beta \epsilon] (1-\alpha)\frac{1}{Z_t}-\alpha (t-t_0)}\ .
\end{equation}
The probability that an individual  at time $t$ has Fisher fitness   $\epsilon$ independently of $t_0$ is 
\begin{equation}
P(\epsilon,t)=\int_0^t dt_0 \rho(\epsilon) e^{\int_{t_0}^t dt\exp[-\beta \epsilon] (1-\alpha)\frac{1}{Z_t}-\alpha (t-t_0)}.\nonumber
\end{equation}
This system reaches steady state in the limit $t\to \infty $ with the equilibrium distribution given by the Fermi-Dirac distribution
  ${P}_F^{\star} (\epsilon)$, and in particular we have 
\begin{equation}
{ P}^{\star}_F(\epsilon)=\frac{1}{\alpha}\rho(\epsilon)\frac{1}{e^{\beta (\epsilon-\mu_F)}+1}
\end{equation}
with  $e^{\beta \mu_F}=\Avg{e^{\beta \epsilon}}\alpha/(1-\alpha)$.
This distribution is the Fermi-Dirac distribution.
The duality between Fermi-Dirac and Bose-Einstein distribution has been also recognized in the framework of evolving network models   \cite{Complex, Fermi} and of models for evolving  ecologies   \cite{Ferretti}.

\section{Evolution of sexual populations}
In sexual populations, the somatic cells of diploid organisms have two homologous copies of each chromosome.
In fact the  genome of each diploid individual  is given by the pairs of chromosomes $A$ and $B$ of  the two haploid   gametes coming  from the father and from the mother of the individual.
Let us suppose that   each gamete  is identified  by $N$ loci indicated with Latin letters $i=1,2,\ldots, N$. If we indicate with $\sigma_i$ the allelic state  at each locus  $i$,  the gamete is characterized by the sequence $\{\sigma\}=(\sigma_1,\sigma_2,\ldots,\sigma_i,\ldots,\sigma_N)$ with  $\sigma_i=1,2,3,4$  indicating respectively the   nucleotides adenine, thymine, cytosine, guanine.
Given this description of the gametes, each individual is characterized by the sequence $\{\sigma^A,\sigma^B\}=[(\sigma_1^A,\sigma_2^B)\ldots(\sigma_i^A,\sigma_i^B), \ldots, (\sigma_N^A,\sigma_N^B)\}$ where  $\sigma_i^{A/B}$ indicates the allelic states in each parental gametes  $A/B$.
The reproductive rate, or Wright fitness, $W(\{\sigma^A,\sigma^B\})$ of  a diploid  individual depends on the  pairs of chromosomes   and can be expressed according to the relation
\begin{equation}
W(\{\sigma^A,\sigma^B\})=e^{-\beta U(\{\sigma^A,\sigma^B\})}
\label{WNE}
\end{equation}
where $U(\{\sigma^A,\sigma^B\})$ is the Fisher fitness of the individual with genotype $\{\sigma^A,\sigma^B\}$ and $\beta$ plays the role of the selective pressure.

The gametic life-cycle describes the information transfer at next generations. Two gametes $A/B$ generate a new individual ({\it fertilization}) and the new individual, if it reaches the reproductive state, carries  the information and gives rise (by {\it  meiosis}) to new gametes $S_1,S_2,\ldots S_n$ with $n=0,1,2\ldots$ which continue the gametic life-cycle.
The process of meiosis is a process of reduction in the genetic information of each diploid individual to generate gametes which have only half of the number of chromosomes.
During meiosis  a  process of recombination can occur with small probability at given locations (recombination hotspots) on the chromosome. When a recombination event occurs, homologous sites on two chromosomes can mesh with one another and may exchange genetic information.

The evolution of diploid populations can be studied by similar techniques as used for the evolution of asexual haploid populations.
In particular we might distinguish between the the dynamics assuming non-overlapping or overlapping generations.

\subsection{Non-overlapping generations}
In the case in which the generations do not overlap we  write the evolutionary dynamics in terms of the probability $\tilde{P}^t(\{\sigma\})$ that a gamete has allelic state $\{\sigma\}$ at generation  $t$.
The evolution of this probability follows the equation
\begin{widetext}
\begin{equation}
\tilde{P}^{t+1}(\{\sigma\})={\mathbf M}_{\{\sigma\}|\{\sigma^A,\sigma^B\}}\left[\frac{e^{-\beta U(\{\sigma^A,\sigma^B\})}\tilde{P}^t(\{\sigma^A\})\tilde{P}^t(\{\sigma^B\})}{Z_t}\right].
\label{evSP}
\end{equation}
\end{widetext}
This equation expresses the evolutionary principles in such a way that evolution favors on one hand the reproduction of high fitness individuals and on the other hand recombination and mutation processes that enhance the variation in the population.  These principles are the origin of the stochastic nature of evolution.
We note here that Eq. $(\ref{evSP})$ is   significantly different from Eq. $(\ref{pno})$. 
In fact Eq. (\ref{evSP}) is a quadratic equation in $\tilde{P}^{t}(\{\sigma\})$ whenever Eq. $(\ref{pno})$ is a linear equation.
The operator ${\mathbf M}_{\{\sigma\}|\{\sigma^A,\sigma^B\}}$ introduced in Eq. $(\ref{evSP})$ indicates the free recombination of genetic material and the mutations occurring when a new gamete is generated.
In particular the operator  ${\mathbf M}_{\{\sigma\}|\{\sigma^A,\sigma^B\}}$  is defined as the average over the probability of free recombination  and mutations $Q(\{\sigma\}|\{\sigma^A,\sigma^B\})$. Therefore  the action of the operator over a generic function $f(\{\sigma^A,\sigma^B\})$ is given by
\begin{widetext}
\begin{equation}
\hspace*{-5mm} {\mathbf M}_{\{\sigma\}|\{\sigma^A,\sigma^B\}}\left[f(\{\sigma^A,\sigma^B\})\right]=\sum_{\{\sigma^A\},\{\sigma^B\}}Q(\{\sigma\}|\{\sigma^A,\sigma^B\})f(\{\sigma^A,\sigma^B\})
\label{M2}
\end{equation}
\end{widetext}
where 
\begin{eqnarray}
Q(\{\sigma\}|{\{\sigma^A,\sigma^B\}})&=&\nonumber \\
&&\hspace*{-35mm}=\prod_{i=1}^N\left[\frac{1-\mu}{2}\delta(\sigma_i,\sigma_i^A)+\frac{1-\mu}{2}\delta(\sigma_i,\sigma_i^B)+\frac{\mu}{4}\pi(\sigma)\right].
\label{Q2}
\end{eqnarray}
The functional  expression of $Q(\{\sigma\}|{\{\sigma^A,\sigma^B\}})$  given by Eq. $(\ref{Q2})$ describe the fact that in sexual populations both, recombination processes and mutations (occurring with rate $\mu$) enhance the variation in the population.
The partition function $Z_t$ in Eq. $(\ref{evSP})$ is given by 
\begin{equation}
Z_t=\sum_{\{\sigma^A\},\{\sigma^B\}} e^{-\beta U(\{\sigma^A,\sigma^B\})} P(\{\sigma^A\})P(\{\sigma^B\})\ .
\label{ZpN}
\end{equation}
Similarly to what happens for asexual evolution, the solution of Eq. $(\ref{evSP})$ is given by a path integral over the phylogenies of the diploid sexual populations.
Nevertheless, the phylogenetic histories associated to one genome are much more complex than in the asexual case.
 In asexual populations,  every individual of the population is the descendent from a linear lineage. In fact every individual has a single progenitor at any given generation.
Instead the phylogenetic history of a single individual in a sexual population is by itself a tree of two parents, four grandparents and so on back in time.

\begin{figure}
\includegraphics[width=70mm, height=90mm]{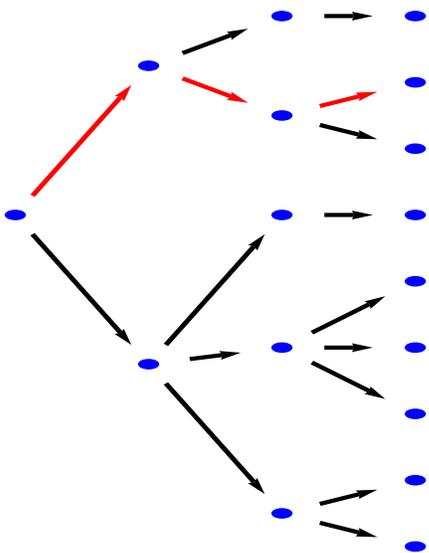}
\caption{(Color online)  The descendant of an individual in an asexual population can be described by a phylogenetic tree.
The history that has given rise to each individual is nevertheless a single path on this tree (indicated in the figure with red arrows). }
\label{Haploidtree.fig}
\end{figure}

\begin{figure}
\includegraphics[width=60mm, height=90mm]{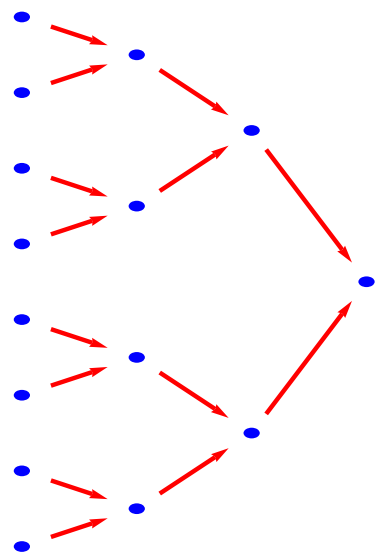}
\caption{(Color online)The phylogenetic  history for an individual  in a sexually reproducing species is the tree of his ancestors.
It is interesting to observe that while the trees of ancestors realistically describes the evolutionary history that has given place to an individual, the global population of a species is not in general decreasing. In fact many of the primordial  ancestors of a given individual in a  species are likely to coincide.}
\label{Diploidtree.fig}
\end{figure}

In Fig. $\ref{Haploidtree.fig}$ we  show the phylogenies of asexual organisms as  linear lineages that combined together form a tree. Instead in Fig. $\ref{Diploidtree.fig}$ we show a single  phylogenetic history of a diploid individual which is by itself a tree. Note that  for ancient generations the  progenitors at the different  leafs of the tree might correspond to the same individual.
We indicate with ${\cal T}(t)$ the phylogenetic tree that starts at generation ${\cal G}(0)$ and ends at generation ${\cal G}(t)$. Each generation ${\cal G}(i)$ is formed by $2^{t-i}$ individuals and the entire tree ${\cal T}(t)$ is formed by ${\cal N}(t)=(2^{t+1}-1)$ individuals.
In the following we  indicate by Greek letters $\alpha, \gamma$ the gametes, represented in the phylogenies represented in  Fig. $\ref{Diploidtree.fig}$ by the links of the tree. Using this nomenclature we can write the path integral solving Eq. $(\ref{evSP})$, i.e.  
\begin{widetext}\begin{equation}
\tilde{P}^{t}(\{\sigma\})=\frac{1}{{{\cal Z}_t}}\sum_{\{\sigma\}\in{{\cal G}(t-1)}}\sum_{\{\sigma\}\in{{\cal G}(t-2)}}\ldots \sum_{\{\sigma\}\in{{\cal G}(0)}}\prod_{\alpha\in {\cal T}(t)}^{{\cal N}(t)}Q(\{\sigma_{\alpha}\}|\{\sigma_{\alpha}^A,\sigma_{\alpha}^B\})e^{-\beta U(\{\sigma^A_{\alpha},\sigma^B_{\alpha}\})}\prod_{\gamma\in {\cal G}(0)}\tilde{P}^0(\{\sigma\}_{\gamma})
\label{pis}
\end{equation}
\end{widetext}
where the normalization constant is given by 
\begin{widetext}\begin{equation}
{{\cal Z}_t}= \sum_{\{\sigma\}\in{{\cal G}(t)}}\sum_{\{\sigma\}\in{{\cal G}(t-1)}}\ldots \sum_{\{\sigma\}\in{{\cal G}(0)}}\prod_{\alpha\in {\cal T}(t)}^{{\cal N}(t)} Q(\{\sigma_{\alpha}\}|\{\sigma_{\alpha}^A,\sigma_{\alpha}^B\})e^{-\beta U(\{\sigma^A_{\alpha},\sigma^B_{\alpha}\})}\prod_{\gamma\in {\cal G}(0)}\tilde{P}^0(\{\sigma\}_{\gamma})\  .
\label{Zs}
\end{equation}
\end{widetext}
In Eqs. $(\ref{pis})$-$(\ref{Zs})$  we indicated with $\{ \sigma_{\alpha}^{A},\sigma_{\alpha}^B \}$ the genotype of the individual  that generates  the  gamete $\alpha$.
We can characterize the operator ${\mathbf M}_{\{\sigma\}|\{\sigma^A,\sigma^B\}}$ by studying how it acts on a base of normalized functions $\phi_n(\{\sigma\})$. In general we will have the relation
\begin{equation}
{\mathbf M}_{\{\sigma\}|\{\sigma^A,\sigma^B\}}\left[e^{-\beta U(\{\sigma^A_{\alpha},\sigma^B_{\alpha}\})}\phi_n(\{\sigma^A\})\phi_m(\{\sigma^B\})\right]=\lambda^r_{mn} \phi_r(\{\sigma\}).
\label{eigp}
\end{equation}
Let us express  the  probability distribution $\tilde{P}(\{\sigma\})$ in the basis of the functions $\phi_n(\{\sigma\})$ according to
\begin{equation}
\tilde{P}(\{\sigma\})=\sum_n D_n(t) \phi_n(\{\sigma\})\ .
\end{equation}
From Eq. $(\ref{evSP})$ we obtain that the system reaches a steady state when  $D_n(t)=D_n^{\star}$ and $Z_t=Z$  with
\begin{eqnarray}
Z \times D_n^{\star}&=&\sum_{rs}\lambda^n_{rs}D_r^{\star} D_s^{\star}\nonumber \\
 Z&=&\sum_n\sum_{r,s} \lambda^n_{rs}D_r^{\star} D_s^{\star}.
 \end{eqnarray}
The detailed analysis of the implications of this solution in specific cases will be performed in subsequent publications.

\subsection{Overlapping generations}

The  probability $P(\{\sigma\})$ that a gamete has an allelic  configuration $\{\sigma\}$, in a continuous time evolution, satisfies the dynamical equation
\begin{eqnarray}
\hspace*{-5mm}\frac{\partial P(\{\sigma\})}{\partial t}&=&{\mathbf M}_{\{\sigma\}|\{\sigma^A,\sigma^B\}}\left[\frac{e^{-\beta U(\{\sigma^A,\sigma^B\})}P(\{\sigma^A\})P(\{\sigma^B\})}{\Avg{W}}\right]\nonumber \\
&&-P(\{\sigma\})
\label{me}
\end{eqnarray}
where   the average fitness is given by 
\begin{equation}
Z_t=\sum_{\{\sigma^A\},\{\sigma^B\}} e^{-\beta U(\{\sigma^A,\sigma^B\})} P(\{\sigma^A\})P(\{\sigma^B\})\ .
\label{Zp}
\end{equation}
The operator ${\mathbf M}_{\{\sigma\}|\{\sigma^A,\sigma^B\}}$ introduced in Eq. $(\ref{me})$ indicates the free recombination of genetic material occurring when a new gamete is generated.
In particular the operator  ${\mathbf M}_{\{\sigma\}|\{\sigma^A,\sigma^B\}}$  is defined as the average over the probability of free recombination   $Q(\{\sigma\}|\{\sigma^A,\sigma^B\})$. Therefore  the action of the operator ${\mathbf M}_{\{\sigma\}|\{\sigma^A,\sigma^B\}}$ over a generic function $f(\{\sigma^A,\sigma^B\})$ is given by Eq. $(\ref{M2})$
with 
$Q(\{\sigma\}|{\{\sigma^A,\sigma^B\}})$ given by Eq. $(\ref{Q2})$.
The equation for sexual evolution is much more complex than the one for asexual evolution since it is non-linear. In order  to solve it
we evaluate, as we did in the model with non-overlapping generations,   the operator ${\mathbf M}_{\{\sigma\}|\{\sigma^A,\sigma^B\}}$ on a basis of normalized functions $\phi_n(\{\sigma\})$.
The operator ${\mathbf M}_{\{\sigma\}|\{\sigma^A,\sigma^B\}}$, will act on any general basis of functions $\phi_n(\{\sigma\})$  according to  Eq. $(\ref{eigp})$.
 Therefore the solution of the sexual evolution is given by 
 \begin{equation}
P(\{\sigma\})=\sum_n C_n(t) \phi_n(\{\sigma\}) 
 \end{equation}
 with the coefficients $C_n(t)$ satisfying the non-linear equation
 \begin{eqnarray}
 \frac{dC_n(t)}{dt}&=&\frac{\sum_{rs}\lambda^n_{rs}C_r(t) C_s(t)}{Z_t}-C_n(t)\nonumber \\
 Z_t&=&\sum_n\sum_{r,s} \lambda^n_{rs} C_r(t)C_s(t).
 \label{DynS2}
 \end{eqnarray}
 The system reaches a steady state only if all the coefficients $C_n$ and the partition function $Z_t$ converge to time-independent values ($C_n^{\star}, Z$, respectively).
 Using $(\ref{DynS2})$ we derive the condition that $C_n^{\star}$ and $Z$  have to satisfy, 
 \begin{eqnarray}
Z \times C_n^{\star}&=&\sum_{rs}\lambda^n_{rs}C_r^{\star} C_s^{\star}\nonumber \\
 Z&=&\sum_n\sum_{r,s} \lambda^n_{rs}C_r^{\star} C_s^{\star}\ .
 \end{eqnarray}
 Although there are many similarities between evolution of sexual and asexual populations, we observe also relevant differences.
 In particular in sexual evolution there is no equivalent of Eq. $(\ref{FD})$, therefore the average reproductive rate is not necessarily increasing in the population.
 
 \subsection{Bose-Einstein condensation in  sexual populations driven by epistatic interactions}
 Epistatic interactions  \cite{Slatkin} between genetic loci are the non-additive terms to the Fisher fitness $U(\{\sigma^A\},\{\sigma^B\})$. Understanding the implications of evolution in presence of epistatic interactions is one of the most promising fields in biological evolution since it might capture essential features for the analysis of the genetic origin of diseases.
 In a recent paper   \cite{Bose_n}  the evolution of sexual populations in presence of pairwise epistatic interactions between genetic loci,  and absence of mutations,  was considered.
In the model studied in \cite{Bose_n} the loci $i=1,2\ldots,N$ are linked in an epistatic network  formed by  $L$ links. 
The epistatic interactions between pairs of loci  have a role in determining the fitness function that can be written  according to the expression
\begin{equation}
W(\{\sigma^A,\sigma^B\})=\prod_{<ij>}e^{-\beta u_{ij}(\sigma_i^A,\sigma_j^A,\sigma_i^B\ ,\sigma_j^B)},
\label{W}
\end{equation}
where the product is extended to all  genetic loci $<i,j>$ linked in the epistatic network.
If the network is locally tree-like,  the general  structure of the solution to the evolutionary equation $(\ref{me})$ is given by  
\begin{equation}
{P}(\{\sigma\})=\sum_{h=1}^H \prod_{<i,j>}b_{ij}^h(\sigma_i,\sigma_j)\ ,
\label{Gtree}
\end{equation}
where $<i,j>$ indicates all the pairs of linked nodes  present in the epistatic network.

In   \cite{Bose_n} the steady states of the evolutionary dynamics of the type as in Eq. $(\ref{Gtree})$ with $H=1$ in presence of epistasis and absence of mutations have been fully characterized.
These steady states are multiple and the distribution of pairs of allelic states is given by a Bose-Einstein distribution.
A Bose-Einstein condensation can occur in which a finite fraction of pairs of genetic loci fixes.

\section{Conclusions}
In this paper we have outlined the similarities between biological evolution of sexual and asexual populations and the formalism of quantum mechanics. Moreover, we  have shown  that the dynamics of biological evolution of non-overlapping generations is  expressed    in terms of path integrals. We have characterized the solution of the equations for evolving sexual and asexual populations finding interesting similarities and significant differences.
Finally we have observed the emergence of quantum statistics in special cases where the population is at stationarity.

In the book {\em What is life?}   \cite{Schroedinger} 
Erwin Schr\"odinger proposed that   a deep relation might exist between biological evolution and quantum mechanics.
Since then this idea has  fascinated biologists and physicists   \cite{Davies, Lloyd, Penrose,McFadden}.
Nevertheless we still lack a firm scientific basis for this suggestive proposal.
We believe that this paper, relating the formalism of biological evolution to the one of quantum mechanics will open new perspectives to  investigate further the relation between biological evolution and quantum mechanics. 
In the future we plan to study evolution in presence of specific fitness functions including   epistasis between genetic loci  and in adaptive networks in which the fitness function changes with time in a non stationary fashion, describing the evolution of species.

\end{document}